# Image Encryption Decryption Using Chaotic Logistic Mapping and DNA Encoding

Sakshi Patel
Dept. of Communication Engineering,
Vellore Institute of Technology,
Tamil Nadu, India
thesakshipatel@gmail.com

Bharath K P
Dept. of Communication Engineering,
Vellore Institute of Technology,
Tamil Nadu, India
bharathkp25@gmail.com

Rajesh Kumar Muthu
Dept. of ECE,
Vellore Institute of Technology,
Tamil Nadu, India
mrajeshkumar@vit.ac.in

*Abstract*— Encryption is a technique used to secure and protect the images from unfair means. Nowadays due to digital world, it is very easy to hack any information which is present in the cloud. Security is an important issue in anyone's life, and encryption plays a vital role to ensure security. There exist many image encryption algorithms that are mainly based on chaotic logistic maps, but there is drawback like small key space and weak security. In this paper we have proposed a method that uses chaotic logistic mapping and DNA encoding to encrypt the image. A 32 bit ASCII private key is used to diffuse the image. The results demonstrated clearly show that encryption algorithm based on chaotic logistic mapping and DNA encoding gives better result than encrypting only with chaotic logistic mapping. The proposed method also takes into account the possible parametric like PSNR and SSIM.

## I. INTRODUCTION

Due to advancement in technology is has become easy to share information like images, data, voice in seconds. As this information is shared over a single band of frequency which can be questioning the security of personal information from the end user side. Therefore the techniques used in order to secure the information plays an important role in maintaining integrity, privacy and authentication of the data we share from unauthorized users. In this present era of technology and digitization, security plays an important role, so encryption is one of the ways to secure our data from hacking. In this paper we will be discussing al algorithms to encrypt images efficiently using different techniques.

Many encryption techniques have been proposed, each of them have their own advantages and disadvantages. Among these there is one algorithm known as chaos based cryptographic algorithm. This algorithm is suggested to be efficient in encrypting images. Chaos algorithm for encryption is considered good, as it provides high speed, reasonable computation, and good security. This system contains some noisy behavior but it is exactly deterministic.
If initial value and its parameters are given then we can generate a confusion matrix. This map is extremely sensitive to initial conditions. Here in this paper we have also used DNA encoding technique that helps to make the encryption far more confusing and random. DNA encoding is a technique to encode the pixel values into a DNA sequence of nucleic acid bases A, T, G, C. We proposed a new approach to encrypt images using chaotic logistic mapping and DNA encoding in order to get secure encrypted image.

Many techniques are presently there for encryption and decryption of multimedia data of 1D, 2D and 3D level. Image encryption techniques are studied frequently in order to meet the demand for real-time information security when data is being transferred over internet. Traditional algorithm i.e., Data Encryption Standard (DES) has many disadvantages like low-level efficiency with large multimedia, used basically for data encryption not for multimedia. The chaotic encryption has being suggested to be fast and highly secured technique for encryption
The various applications of this method of image encryption are:
1. Medical imaging system
2. Confidential video conferencing
3. Military image databases
4. Online personal photograph album
5. Cable TV, etc

In the proposed image encryption scheme, an external secret key of 32-bit, a chaotic logistic map and DNA encoding scheme are employed. The initial conditions for the logistic map are derived using the external secret key by applying logical operations all its bits and further encoding them to DNA sequence.
Further, in the proposed encryption process, the input image and key image will be operated to give the final encrypted image.

## II. LITERATURE SURVEY

In this section we will be discussing the various existing methods present for image encryption and decryption.

A New Chaotic Algorithm for Image Encryption by Xin Zhanga , Weibin Chenb, have discussed a new image encryption algorithm, where he is using Henon chaotic maps for secure image encryption for safe transfer. Gray values present in the image are distributed in a random fashion. In this Henon chaotic map's initial value and key image are used as secret key.

Jin-mei Liu, Qiang Qu, "Cryptanalysis of a substitution–diffusion based image cipher using chaotic standard and logistic map", in this paper they have used standard map, which is a used in image encryption schemes.

Hong, Lianxi, and Chuanmu Li. "A novel color image encryption approach based on multi-chaotic system", here multi-chaotic system is discussed for color image encryption.

Honglei, Yu, Wu Guang-shou, "The compounded chaotic sequence research in image encryption algorithm", in this research they have combinedly used some of the unlimited folded map, which are 3D- Baker map, Henon map and logistic map, for encryption of images. But the future scope in this paper was; the histogram plotted after this strong encryption was not uniformly distributed.

Zhang D., Gu Q., Pan Y. and Zhang X. "Discrete Chaotic Encryption and Decryption of Digital Images", this paper scrambled the row and column for each pixel in the input image using two 1-D discrete Chebyshev chaotic sequences.

Al-Najjar, Hazem Mohammad, Asem Mohammad AL-Najjar, K. S. A. Arar "Image Encryption Algorithm Based on Logistic Map and Pixel Mapping Table", This research work have used two methods for encryption: Pixel Mapping Table (PMT) and logistic mapping. Firstly PMT was used to confuse and increase uncertainty in the original image. Then rows and columns replacement was done. Lastly XOR logical operation was applied with random vector generated using logistic map.

## III. METHODOLOGY

In this section we aim at the techniques used for image encryption and decryption.

### A. Encryption

Encryption is a technique to secure data like; multimedia or confidential documents by a process of encoding it in such a way that only allowed parties can have access to it. These authorized parties can only have access to these files with a decryption key which will be shared by the transmitter. We will be using symmetric and private key to encode the input image.

Symmetric Keys: In this scheme, encryption and decryption keys are same. Transmitter and receiver should have same keys to have a secure transmission of data.

Private Keys: In private key encryption schemes, the encryption key is known only to the transmitter and to the receiver to encrypt messages.

The different types of techniques used in this paper for encryption of images are:

#### 1. Chaotic Logistic Map:

The chaotic theory is the mathematics considering dynamic behavior of the natural and artificial systems which are sensitive to the initial conditions like weather, climate and traffic on road. It can be analyzed using chaotic mathematical model or can also be done using recurrence plots and Poincare maps. Chaos theory is used for emerging technologies such as neurology, cardiology, control and circuit theory, weather prediction, etc. Chaos -"when the present determine the future but the approximate present cannot approximately determine the future". In chaos even small change in initial conditions can lead to totally uncorrelated sequence. It is been said and proved that chaos function can be used for encryption giving good results. Logistic function is one of the chaos functions varying with initial conditions with high sensitivity and generated non periodic pseudo random sequence and it will be entirely unpredictable if proper choice of bifurcation parameter 'r' has been taken in to consideration. Implementing image encryption by using chaotic theory is simple, computationally faster and impregnable.

Logistic map is the function which is given by the equation 1 which produces 2-D non periodic chaotic sequences $\{p_i\}$, where $p_i$ is lies between 0 and 1 and is random in nature.

$$p(n+1) = u\, p(n)\, (1-p(n)) \qquad (1)$$

Here u is known as bifurcation parameter having range from 0 to 4.
$p(0)$ is the initial value $0 < p < 1$ and $\{p_1, p_2, p_3 \ldots p_n\}$ sequence elements are generated as per the equation 1.
As per the research results, system will have chaotic nature when 'u' ranges from $3.56994 < u \leq 4$.
Fig.1 shows the bifurcation diagram of logistic map proving that this map is mostly chaotic when values are nearly 4.

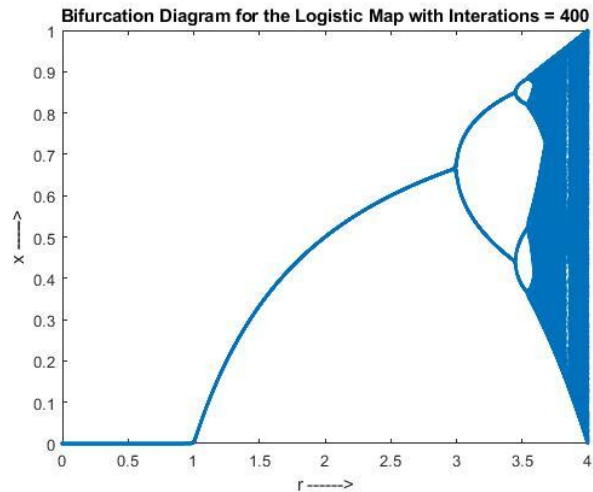

Fig.1 Chaotic behavior of the logistic map

#### 2. DNA Encoding:

DNA encoding technique is used for encoding and decoding of images. As we know in medical, DNA sequences have four nucleic acid bases, which are T (thymine), C (cytosine), A (adenine), G (guanine). Here, C and G are complementary, and T and A are complementary to each other. Now if we come to binary, 1 and 0 are complementary of each other. In this paper we are using A, T, C, G denoted as 00,11,01,10 respectively. If we represent 8 bit gray image in a DNA sequence then each pixel will have length of 4. Take an example: first pixel of the image is 173, now converting it into binary stream as [10101101], then using the above DNA encoding rule in order to encode obtained stream of bits we will get [GGTC]. Where 00=A, 11=T, 01=C, 10=G, and again from DNA sequence we can generate binary sequence easily. From this example we can conclude that 1 pixel having gray value 173 can be represented in DNA encoded bits having sequence [GGTC]. Now from the four DNA nucleic acids we can have 24 different types of combinations. But however we can have only eight combinations which are following the principle of complementarily. These rules are summarized in Table 1.

Table 1: DNA rules

| 1 | 2 | 3 | 4 | 5 | 6 | 7 | 8 |
|---|---|---|---|---|---|---|---|
| 00-A | 00-A | 00-C | 00-C | 00-G | 00-G | 00-T | 00-T |
| 01-C | 01-G | 01-A | 01-T | 01-A | 01-T | 01-C | 01-G |
| 10-G | 10-C | 10-T | 10-A | 10-T | 10-A | 10-G | 10-C |
| 11-T | 11-T | 11-G | 11-G | 11-C | 11-C | 11-A | 11-A |

Size of original gray scale image is M*N, after DNA encoding the size will be increased to 4*M*N.

$$Io=\{I1,I2,x3,\ldots\ldots,IMN\} \quad (2)$$

$$I(encoding)=\{I1,I2,I3,\ldots\ldots,I4MN\} \quad (3)$$
$$Ii = \{A,T,G,C\}$$

*B. Image Encryption Algorithm*

Over all block diagram of image encryption is shown in figure 2.

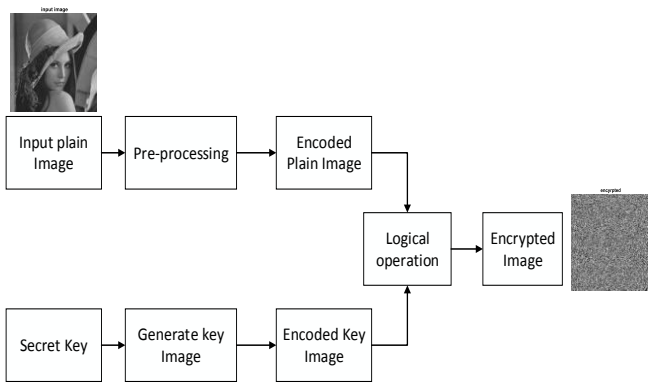

Fig. 2 Encryption Block Diagram

The algorithm for image encryption is:

Step 1. Resize the input image and convert it into gray scale.

Step 2. Use 32 bit ASCII number as a key to encode the input image. Convert hexadecimal key to 128 bit binary. Selecting any 32 bits and divide them into 4 parts having 8 bit each. Now apply 'xor' operation on these bits and convert the answer into decimal.
This decimal number will be used as the initial input to chaotic map.
Step 3. Apply chaotic logistic map, taking initial input from the previous step, and 'r' value to be 3.99999 (most chaotic).

Step 4. Apply 8 rules of the DNA encoding to convert the input image into DNA encoded image.

Step 5. Take a key image and repeat the same step as above to encode it to get a cipher image.

Step 6. Now as the cipher image and the encoded input image are DNA encoded, we will 'xor' both the images using rules to get an encrypted DNA encoded image.

Step 7. Now as the output image from step 6 is DNA encoded, we will be converting each pixcel to decimal value using rules to get the final encrypted image.

*C. Image Decryption Algorithm:*

Over all block diagram of image encryption is shown in figure 3.

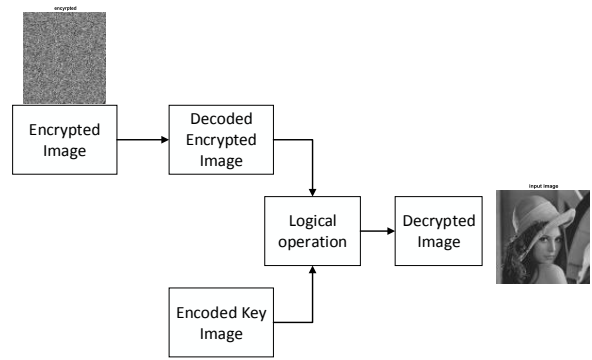

Fig.3 Decryption block diagram

Take the encrypted image and the cipher image; apply all the steps in the reverse direction as done in the encrypted algorithm. The results obtained will the decrypted image

IV. RESULTs

*A. Image Encryption:*

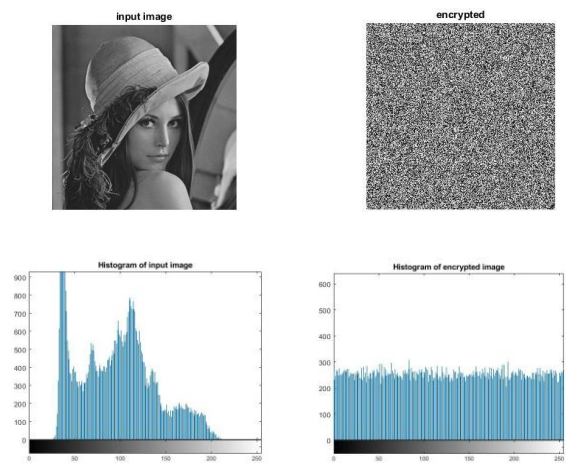

Fig.4 Input image with histogram, encrypted image with histogram

## B. Image Decryption:

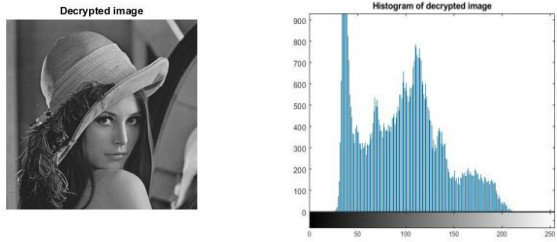

Fig.5 Decrypted image with histogram

## C. Parametric Measurements:

1. *Peak to Signal to Noise Ratio (PSNR):*

PSNR represents the peak error in the image. This parameter should be large, as it represents the ratio of signal power to noise power, noise power should be minimum.

2. *Structural Similarity Index (SSIM):*

This parameter is used to measure the pixel similarity between input image and the decrypted image.

3. *Encrypted Image Histogram:*

In Fig.6 it is shown that the histogram on the left side of the figure is obtained when only chaotic logistic map is applied on the input image to get encrypted image. Histogram on the right side is obtained when chaotic logistic map and DNA encoding is applied with 32 bit ASCII key on the input image to get encrypted image.

The histogram on left side is almost similar to the histogram of the input image, which tells that encryption of the image is not done properly and the image can be predicted and hacked by any one while transferring image from transmitter to the receiver. But right side histogram is flat, which tells that the encryption is done well to an extent and hacker will have problem in predicting it.

Table 2 shows the different parametric measurements to analyze the result.

Table 2: Parametric Measures

| Matrices | Chaotic Logistic Map | Chaotic Logistic Map and DNA Encoding |
|---|---|---|
| **PSNR** | 52.8641 | 55.1309 |
| **SSIM** | 0.9899 | 0.9987 |
| **Histogram** | 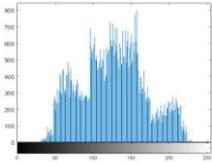 | 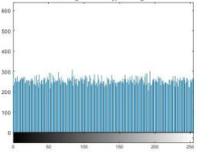 |

## V. CONCLUSION

In this paper we proposed an efficient algorithm for image encryption. First, preprocessing on input image is done to convert it to gray scale image. Then 32 bit ASCII key was used in order to initialize the chaotic map. This chaotic map will create a confusion matrix, this matrix will be encoded using DNA encoding. Cipher key was also created as the input image was encoded. Atlast the encoded input image and the cipher key was 'xor' with each other to create the encrypted image. The results obtained from the encryption and decryption algorithm are shown and analyzed using different parametric measurements.

The future scope of this work can be done while generating the cipher image using key.